\documentclass[final,3p,times,twocolumn]{elsarticle}

\usepackage{graphics}
\usepackage{graphicx,subfigure}
\usepackage{amssymb}
\usepackage{amsmath,textcomp,array}
\usepackage{multirow}

\journal{Astronomy and Computing}

\begin{document}

\begin{frontmatter}



\title{An Ultra Fast Image Generator ({\sc UFig}) for wide-field astronomy}
\author{Joel Berg\'e\corref{cor1}}
\author{Lukas Gamper, Alexandre R\'efr\'egier and Adam Amara}
\address{ETH Zurich, Department of Physics, Wolfgang Pauli Strasse 27, 8093 Zurich, Switzerland}
\cortext[cor1]{jberge@phys.ethz.ch}

\author{}

\address{}

\begin{abstract}

Simulated wide-field images are becoming an important part of observational astronomy, either to prepare for new surveys or to test measurement methods. In order to efficiently explore vast parameter spaces, the computational speed of simulation codes is a central requirement to their implementation.
We introduce the Ultra Fast Image Generator ({\sc UFig}) which aims to bring wide-field imaging simulations to the current limits of computational capabilities. We achieve this goal through: (1) models of galaxies, stars and observational conditions, which, while simple, capture the key features necessary for realistic simulations, and (2) state-of-the-art computational and implementation optimizations. We present the performances of {\sc UFig} and show that it is faster than existing public simulation codes by several orders of magnitude. It allows us to produce images more quickly than {\sc SExtractor} needs to analyze them. For instance, it can simulate a typical 0.25 deg$^2$ Subaru SuprimeCam image (10k$\times$8k pixels) with a 5-$\sigma$ limiting magnitude of ${\rm R}=26$ in 30 seconds on a laptop, yielding an average simulation time for a galaxy of 30$\mu$s. 
This code is complementary to end-to-end simulation codes and can be used as a fast, central component of observational methods relying on simulations.
\end{abstract}

\begin{keyword}
Simulations \sep Wide-field imaging \sep Computational speed


\end{keyword}

\end{frontmatter}


\section{Introduction}

Image simulations are becoming ubiquitous in observational astronomy. They are intensively used in topics as diverse as extragalactic astrophysics (with public codes like {\sc Skymaker} \cite{skymaker}, {\sc SImage} \cite{simage}, {\sc Shera} \cite{mandelbaum12}, or the simulation package developed by the Large Synoptic Survey Telescope (LSST) team \cite{lsst}), CMB analysis (e.g. \cite{delabrouille12}), or Supernovae (e.g. \cite{bernstein12,perrett10}). Simulations are mainly used to forecast the results of an observation strategy and to test measurement methods. Examples are given by the LSST simulations \cite{lsst,chang12a,chang12b}, the public Shear Testing Program (STEP) and Gravitational Lensing Accuracy Testing (GREAT) simulations \cite{step1,step2,great08,great10}, as well as individual works (e.g. \cite{berge12,kacprzak12}).

In this work, we focus on simulations used for extragalactic wide-field astronomy. Depending on their aim, such simulations will either be minimalist, like the GREAT simulations which simulated simplistic individual galaxies  on a mesh, or include most of, or all relevant cosmology, astrophysics, atmosphere physics and telescope characterization (LSST). The STEP project (using {\sc Skymaker} and {\sc SImage}) took an intermediate approach, where simulations look realistic, but without focusing much on any particular physics or observational apparatus.

Depending on their emphasis, the speed of simulation codes can vary greatly, from hours to days to simulate a quarter square degree ground-based image.
However, if a simulation package is to be used as an integrated part of a method and pipeline calibration process, its speed becomes a driving parameter: a fast simulation code, used to calibrate an external measurement method, allows one to efficiently explore a larger parameter space for the measurement method or survey.

The aim of this paper is to introduce the Ultra Fast Image Generator ({\sc UFig}), a fast simulation C++ code able to simulate realistic images on a timescale comparable to that needed by {\sc SExtractor} \cite{bertin96} to analyze similar images (i.e., less than one minute for a 0.25 sq. deg. image); since it is widely used in astronomy and is well optimized, we take {\sc SExtractor} as a reference. 
 To this end, we adopt simple, yet realistic, models of galaxies, stars and observation conditions, that allow us to minimize the computation load. We also bring our code to the current computation limits by highly optimizing our implementation through an efficient use of random number generators, parallelization and vectorization.
 This fast code is thus complementary to end-to-end simulation packages aimed at detailed modeling of observational effects.

Although the code may be used to simulate an image from various ground-based facilities and observation conditions, in this paper we use as a practical test case the simulation of a typical Subaru SuprimeCam \cite{miyazaki02} coadded 10k$\times$8k pixels image, processed from four 450-seconds exposures, with a 5-$\sigma$ (extended) limiting magnitude of ${\rm Rc}=26$, unless otherwise stated.

Section \ref{sect_model} summarizes the model that we use for galaxies, stars and noise. Further details about it can be found in the appendices. Section \ref{sect_req} motivates the strategy used to optimize the simulation of realistic images, and Section \ref{sect_implementation} describes the implementation and the optimizations we use on the computational part of the problem; in particular, we present how we optimize random numbers generation and implement multithreading. Section \ref{sect_imcon} explores the consistency of the {\sc UFig} simulations with real images. Section \ref{sect_perfs} shows the performances of our code; in particular, we show in this section how the execution time depends on the image's size and on the exposure time. We conclude in Sect. \ref{sect_ccl}. Further details about {\sc UFig}, including examples and information about the distribution of the code, can be found at http://www.astro.ethz.ch/refregier/research/Software/ufig.

\section{Model} \label{sect_model}

The simulations are based on a simple, yet realistic, modeling of galaxies, stars and noise: the models are summarized in this section. The appendices expand on the astrophysics and the methods used to assess our models.

\subsection{Galaxies}

We assume that galaxy profiles are well described by the Sersic profile \cite{sersic68}
\begin{equation} \label{eq_sersic}
I(r)=I(r_{50})\exp\left( -k\left[\left(\frac{r}{r_{50}}\right)^{1/n}-1\right] \right),
\end{equation}
where $r_{50}$ is the radius of the circle enclosing 50\% of the total flux of the galaxy, $n$ is the Sersic index, and $k$ is a constant satisfying the equation $2\gamma(2n,k)=\Gamma(2n)$, where $\gamma(x,k)$ is the lower incomplete gamma-function, and $\Gamma(x)$ is the gamma-function.

We find the probability density function (p.d.f) of Sersic indices for bright galaxies (magnitude less than 20) to be well represented by $f(n)=\exp(\mathcal{N}(0.3,0.5)+\mathcal{N}(1.6,0.4))+0.24$, where $\mathcal{N}(\mu,\sigma^2)$ represents the normal distribution of mean $\mu$ and variance $\sigma^2$ (see \ref{app_gals}).
For faint galaxies (magnitude bigger than 20), we describe it by $f(n)=\exp(\mathcal{N}(0.2,1))+0.2$.

We parametrize the magnitude distribution of galaxies with a polynomial of the form $\log_{10}(N<{\rm mag})=\sum_i a_i {\rm mag}^i$.
Table \ref{tab_galscounts} summarizes the coefficients $a_i$ for different filters.

Finally, we account for the galaxies intrinsic ellipticity distribution with a 2D Gaussian (of both components of the ellipticity) of width $\sigma_1=\sigma_2=0.15$.

\subsection{Stars and Point Spread Function}

We use a Moffat profile to account for the Point Spread Function (PSF).
The (circular) Moffat profile is defined as \cite{moffat69}:
\begin{equation} \label{eq_moffat1}
I(r)=I_0 \left[1+\left(\frac{r}{\alpha}\right)^2 \right]^{-\beta},
\end{equation}
where $I_0$ is the value at the origin ($r=0$), and $\alpha$ and $\beta$ are scale parameters depending on the observation's conditions. The width of the profile, $\alpha$, is related to its FWHM and to its half-light radius $r_{50}$.

\subsection{Noise}

We finally account for noise in the image. We first add Poisson noise for galaxies and stars. We then add the noise from various sources, like the sky brightness, the readout noise and the errors arising during the data processing, such as flat-field inaccuracies. Following e.g. \cite{grazian04,meneghetti08}, we define it as a Gaussian random deviate with zero mean and standard deviation given by Eq. (\ref{eq_fullnoise}).
We finally correlate the noise with a Lanczos resampling \cite{duchon79}.

\section{Code requirements} \label{sect_req}

\subsection{Requirements}

The main requirement of our simulation pipeline is that it must be fast, while providing realistic images. Since {\sc SExtractor} has become the reference software for wide-field image analysis and is computationally efficient, we use its execution time on a given image as our unit of time. In that sense, we want the running time of the cycle simulation creation ({\sc UFig}) -- simulation analysis ({\sc SExtractor}) not to be dominated by the execution of UFig, hence setting the requirement that {\sc UFig} is not slower than {\sc SExtractor}.
For instance, {\sc UFig} should simulate a typical Subaru SuprimeCam image of 0.25 sq. deg. (made up of approximately 10,000$\times$8,000 pixels) with a 5-$\sigma$ limiting magnitude of ${\rm Rc=26}$ under one minute.

To meet this goal, we must define the most efficient way to draw galaxies and stars, their generation being the most expensive task of a simulation code.

\subsection{Pixel-based or photon-based?}

We can think of two ways to simulate 2D objects such as galaxies and stars in an image from a given light distribution: (1) pixel-based and (2) photon-based.
 
In the former case, an analytic description of the object is pixelized. The description can be a simple profile, which is simply pixelized by taking its value at the center of pixels, or by integrating it in pixels, or it can be more complex, as in a shapelets model \cite{shapelets1,shapelets3}. Public simulation packages like {\sc Skymaker}, {\sc SImage} or {\sc Shera} rely on this principle. 

In the second approach, an analytical description of the object is considered as the distribution of the photons that make it up. Photons are then drawn individually to make up the object. This approach is used e.g. by the LSST Simulation group \cite{lsst}.

The choice of the algorithm is determined by the total number of operations needed to create all the galaxies and all the stars of the simulation. These are described below.

\subsubsection{Number of operations for one elementary building block}

\paragraph{Pixel-based approach:} In this case, the elementary building block of an object is one pixel. To simulate one pixel, we first have to draw its value from the analytical description of the object. The value of the PSF for that pixel is drawn in the same way, from an analytical description of the PSF shape. An object should be made on a refined grid (i.e., using pixels smaller than those of the final simulations) in order to minimize approximations of the profile at the center of pixels; analytically integrating over pixels allows one to get rid of those approximations, but at the price of a more complex implementation.
Then, Poisson noise must be applied to the pixel. Finally, the object, once created, is convolved with the PSF. This last step, albeit optimized by using FFTs, is computationally expensive, even more if the object is refined so that numerical errors are minimized. 

\paragraph{Photon-based approach:} In this case, the elementary building block of an object is one photon. To simulate one photon, we draw its position from the analytical description of the object, seen as a distribution function. The effect of the PSF is simply to displace the photon; therefore, we just have to draw a random displacement from the analytical description of the PSF, and apply it to the position of the photon. Contrary to the pixel-based approach, no complex task (such as a convolution) has to be done at all. Finally, since photons are drawn individually, Poisson noise emerges naturally, with no need to add it eventually.

Therefore, less operations are necessary to simulate a photon than to simulate a pixel. In the next subsection, we consider the number of photons and the number of pixels that we need to simulate to make an extragalactic Subaru SuprimeCam-like image, before concluding on what approach we choose.

\subsubsection{Number of photons vs number of pixels}

\begin{figure} 
\includegraphics[width=8cm]{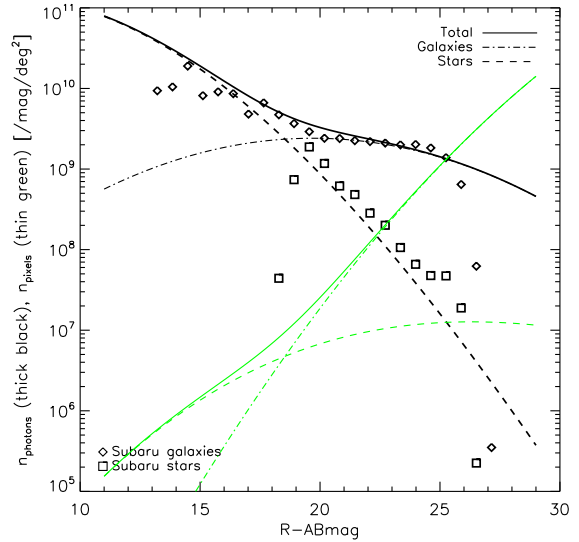}%
\caption{\label{fig_nphotmag} Number of photons (thick black) and number of pixels (thin green) sampled per magnitude per square degree. Dash-dot lines represent the contribution of galaxies, dashed lines show that of stars, and solid lines show the total. Diamonds and squares are photon counts from galaxies and stars from a typical Subaru-SuprimeCam image in Rc-band.}
\end{figure}

The number of photons coming from galaxies per square degree and per magnitude can be computed from the magnitude distribution of galaxies (Eq. \ref{eq_magdist}) and the relation between the number of photons and the magnitude of a galaxy (Eq. \ref{eq_nphotons}). A similar approach allows us to estimate the number of photons coming from stars. Integrating those functions, it is easy to estimate the number of photons required for a photon-by-photon simulation.
Similarly, assuming that an average galaxy is contained in a 21$\times$21 postage stamp, and that we resample it by a factor of 5 when simulating it\footnote{We find that a factor 5 refinement is a bare minimum, and as such, is a lower bound on what can be used. The bigger this factor, the more precise and expensive the simulation.}, we can estimate the number of sampled pixels per square degree and per magnitude in a pixel-by-pixel approach.  The same exercise can be done for stars, assuming that the average postage stamp size is 61$\times$61 pixels (stars are on average brighter and more spread out than galaxies).
Fig. \ref{fig_nphotmag} shows the number of photons per square degree and per magnitude (black thick lines) and the number of (resampled by a factor 5) pixels required per square degree and per magnitude (green thin lines). In both cases, dash lines correspond to stars, dash-dot lines correspond to galaxies, and solid lines show the total number of photons and pixels. The symbols show photon counts from a typical Subaru image: squares correspond to stars and diamonds to galaxies. Note that the star-galaxy separation, performed with {\sc SExtractor}, confuses stars and galaxies at magnitudes less than 18, corresponding to saturated objects. Nevertheless, the total number of photons per square degree and magnitude agrees extremely well with our expectations (black lines).

The number of photons per square degree and per magnitude from galaxies remains fairly constant with magnitude, meaning that adding faint galaxies does not affect the execution time when simulating galaxies in a photon-based approach. On the other hand, photons from stars largely dominate the total number of photons at low magnitudes, and become highly subdominant for intermediate and high magnitudes. Thus, in a photon-based approach, the execution time will be dominated by stars generation.
The same comparison between stars and galaxies can be done for the number of pixels: stars dominate at low magnitude, before becoming largely subdominant. Contrary to the number of photons, the number of pixels from galaxies per square degree and per magnitude increases linearly with magnitude, meaning that going deeper significantly affects the execution time. In the pixel-by-pixel approach, the execution time is dominated by galaxies.

Table \ref{tab_ppn} gives the number of photons and pixels that one needs to sample to simulate a 0.25 deg$^2$ image with an 450 seconds exposure time, from magnitude 12 to magnitude 29. These numbers confirm the conclusions from Fig. \ref{fig_nphotmag}: the number of photons is highly dominated by stars' photons, and the number of pixels is dominated by galaxies. Furthermore, the number of photons needed to simulate galaxies is of the same order than the needed number of pixels.

\begin{table}
\caption{\label{tab_ppn} Number of photons and pixels sampled for one {\sc UFig} simulation.}
\begin{tabular}{ccc}
\hline
\hline
 & Number of photons & Number of pixels \\
 \hline
 Galaxies & 7$\times10^9$ & $5.6\times10^9$ \\
 Stars & $5.1\times10^{10}$ & $3.4\times10^8$ \\
 Total & $5.8\times10^{10}$ & $6\times10^9$\\
\hline
\end{tabular}
\end{table}

\subsubsection{Conclusion}

The photon-based approach requires less operations per elementary building block (photons), than needed by the pixel-based approach. 
Furthermore, many more photons than pixels are needed to simulate stars.
On the other hand, the numbers of photons or pixels to be sampled in either approach to simulate galaxies are similar. Note that according to Fig. \ref{fig_nphotmag}, we would have to simulate less pixels than photons, were we to simulate galaxies up to magnitude ${\rm Rc} \approx 26$ (corresponding to the Subaru telescope limiting magnitude); however, we need to take into account fainter galaxies, which affect the image's noise, and we choose ${\rm Rc} = 29$ as our higher magnitude.

For these reasons, we have decided to adopt a hybrid approach: we simulate galaxies with a photon-based approach, and stars with a pixel-based approach.

\section{Implementation} \label{sect_implementation}

\subsection{Galaxies}

We simulate a circular Sersic galaxy by sampling the radial position $r$ of its $N_\Phi$ photons (Eq. \ref{eq_nphotons} links a galaxy's magnitude and its number of photons) from a $\gamma$-distributed random variable $Y$ ($r=Y^nr_{50}/k^n$), and their angular positions from a uniform distribution between $0$ and $2\pi$ (see \ref{app_gals}). Then, we transform the galaxy so that it becomes elliptical with the desired ellipticity, as shown in the appendix.

Finally, the galaxy is pixelized by truncating the coordinates of the photons' positions. Note that we truncate, and not round, the coordinates because a pixel can be seen as a bucket (e.g., $x=2.8$ corresponds therefore to pixel number 2).

\subsection{PSF-induced photon displacement and stars}

The effect of the PSF on a incident photon is to displace it, the corresponding displacement following a p.d.f defined by the PSF profile. 
This can be seen by considering stars as point-like sources, which would appear as a Dirac function in the absence of a PSF; their observed shape (the PSF itself) is therefore the consequence of the displacement of the photons making up the oncoming Dirac function.
The displacement ${\rm d}X$ due to the PSF is then obtained by uniformally sampling the Moffat profile's cumulative distribution function (c.d.f) ${\rm d}X=\alpha \sqrt{\left[cdf(Y)-1\right]^{\frac{1}{1-\beta}}-1}$ (see \ref{app_stars}). We use this technique to convolve galaxies with the PSF.

As shown in Sect. \ref{sect_req}, stars account for most of the photons to be simulated, hence we create them with a pixel-based approach. To this end, we integrate the Moffat profile numerically in pixels with a 7th order Legendre-Gauss quadrature rule \cite{gausslegendre}, and we perturb each pixel's value with a Poisson deviate. We integrate the profile from the center outwards, until the probability of detecting one photon in a pixel is less than one percent. While much faster, this method is statistically equivalent to drawing photons one by one. Furthermore, as opposed to a pixel-based approach to galaxy generation, creating stars pixel by pixel does not involve expensive numerical convolution with the PSF, nor is it impacted by potential numerical errors coming from the convolution.

\subsection{Optimizations}

\subsubsection{Random number generation}

Operations such as drawing the position of galaxies and stars, drawing galaxies' photons, or generating the background noise, imply heavy use of random number generators.

To enable the creation of the $\approx 3\times10^{10}$ random numbers needed to simulate galaxies\footnote{This number corresponds to four times the number of photons that must be generated (see Table \ref{tab_ppn}), since a photon requires four random numbers: position with respect to the galaxy's center (radial and angular) and displacement due to the PSF (radial and angular).}, we implemented the lagged Fibonacci generator $x_i = (x_{i-21034} + x_{i-44497}) \,\, {\rm mod} \,\, 2^{32}$ \cite{brent92,brent94}, that we initialize with the mersenne twister generator implemented in the boost::mt19937  random number generator \cite{matsumoto98}. 
The lagged Fibonacci generator is buffered, meaning that we generate a full lag of 44497 numbers at once instead of generating them only when needed. Finally, another advantage of the lagged Fibonacci generator is that it allows us to use vectorization (see below) to generate the lag. We tested our generator with the Test U01 test suite, which consists of 160 tests \cite{lecuyer07}: all tests were passed with a p-value inside the range [$10^{-4}$,$1-10^{-4}$]; tests with a p-value outside of this range would be considered as failures. 

\subsubsection{Parallelization}

We start by generating a catalog of stars and galaxies, where astrometry, photometry and shape information is stored. This task is done with an openmp loop. 

Then, we implement multithreading in the following way. Galaxies are sorted by position (each thread thus dealing with a well defined region of the full image), in such a way that each thread gets the same number of photons; in this way, all threads run in the same time (a very bright galaxy does not impair the speed of a given thread). Stars are sorted by position, by making sure that each thread gets the same number of stars to generate. 
Since galaxies and stars are sorted by position, each thread works safely on its own part of the simulation, completely independently from the other threads. Therefore, we do not need critical section (openmp locks) to write to the global array making up the image. Locks are further avoided by forcing each thread to have its own set of random number generators (independent of other threads' random number generators).

The remaining tasks (noise generation, magnitude rescaling, image resampling) are parallelized using openmp.

\subsubsection{Other optimizations}

\paragraph{Approximation of functions}

We use linear interpolations to common functions such as trigonometric functions or $\Gamma$ function. This significantly speeds up our calculations.

\paragraph{Vectorization}

Streaming SIND Extensions (SSE) allows us to perform four floating point calculations at once: we use it for galaxy generation, noise generation and image resampling. We checked that using floating point values instead of double points value does not impact the precision.

\section{Quality assurance} \label{sect_imcon}

\begin{figure*}
\centering
\includegraphics[width=12cm,angle=0,trim=3cm 2cm 2cm 10cm]{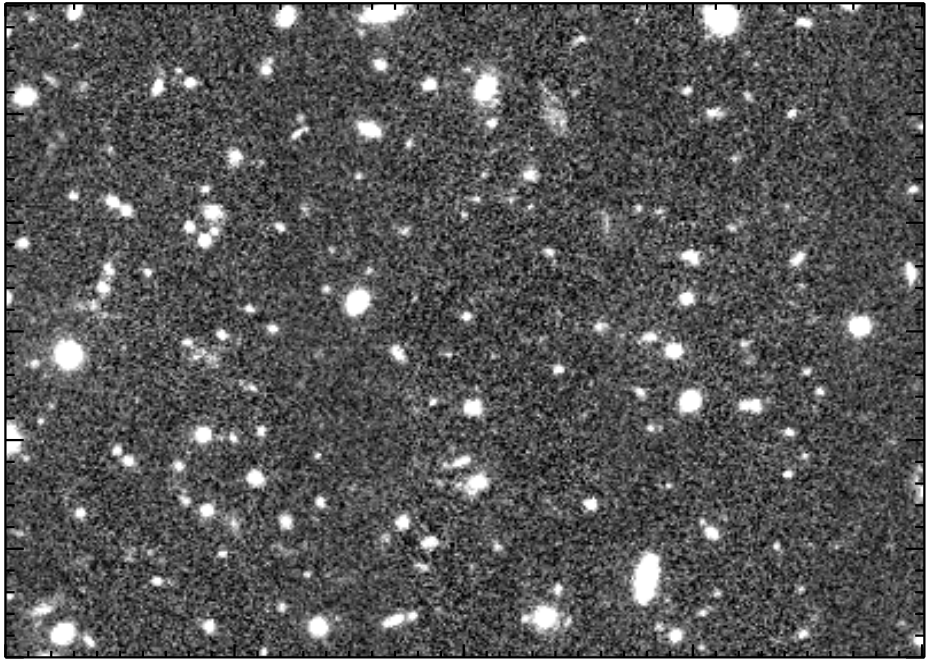}
\includegraphics[width=12cm,angle=0,trim=3cm 2cm 2cm 4cm]{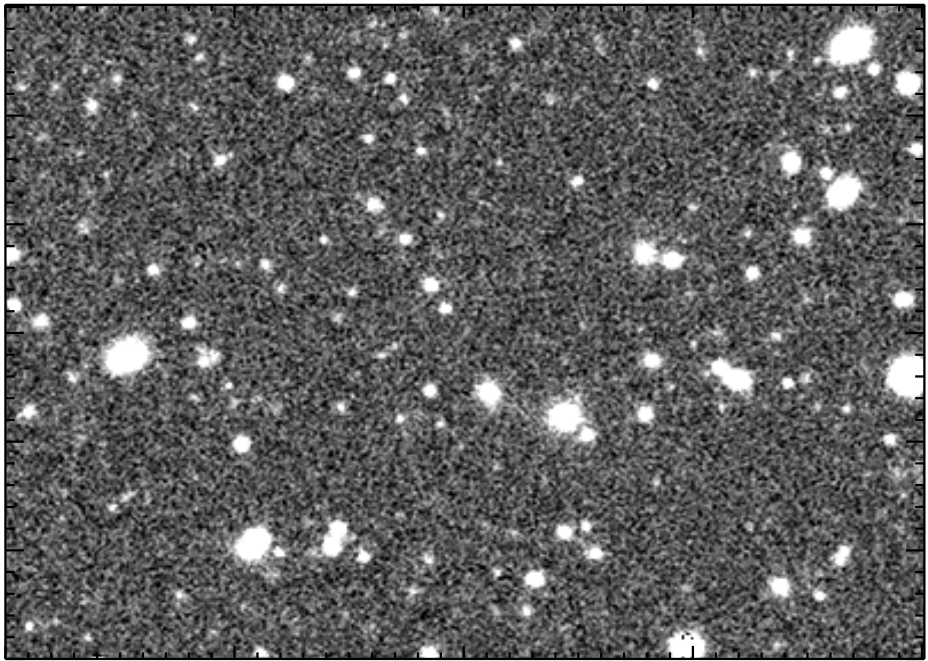}
\caption{\small Top: typical Subaru image. Bottom: {\sc UFig} simulation. Both patches have an area of 400$\times$300 pixels$^2$, or 1.3$\times$1 arcmin$^2$. The dynamic scales are the same for both panels.} \label{fig_compim}
\end{figure*}

Fig. \ref{fig_compim} compares a patch of a {\sc UFig} simulation (lower panel) with a patch of a typical Subaru image (upper panel). Both patches are of the same size, and the dynamic scale is the same in both panels. Visually, the shape and size of galaxies are well rendered by our simulations. Moreover, the granularity and spatial correlation of the background is comparable to that of the real image. We simulate correlated noise by resampling the original simulation, whose background noise is uncorrelated, normally distributed, with a Lanczos resampling. This resampling is fast (see Sect \ref{sect_perfs}), and allows us to mimic the resampling done in the processing of real data.

\begin{figure}
\centering
\includegraphics[width=8cm,angle=0]{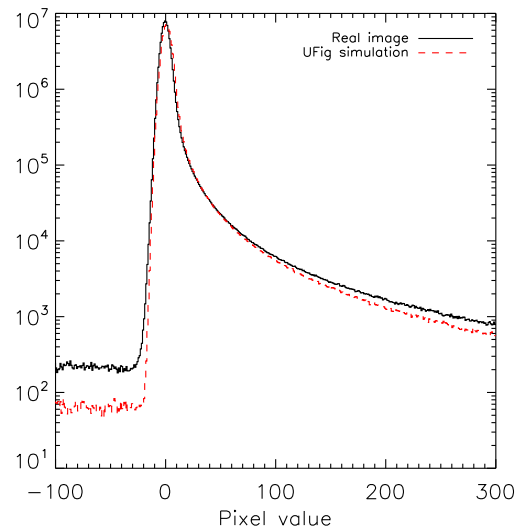}
\caption{\small Pixels distribution. The solid black line is for a typical Subaru image, and the dashed red line is for a {\sc UFig} simulation.} \label{fig_pixdist}
\end{figure}

In Fig. \ref{fig_pixdist}, we show the distribution of pixels in the real image (black solid line) and in the simulation (red dashed line). The distributions agree well, especially for values near zero, corresponding to background pixels, and for high-value pixels, corresponding to bright objects. Around zero, the distributions are well rendered by Gaussian distributions, hence confirming that generating the background noise from a Gaussian deviate is correct.
The flat distribution of negative values for the real image are created when processing raw data single exposures into a resampled, coadded image. Although they are not reproduced with the original Gaussian noise model used in {\sc UFig} (not shown on the figure), we have checked that they appear when we simulate raw data single exposures that we process to obtain a final coadded image. This confirms that they are indeed due to the data processing. Resampling the (uncorrelated Gaussian noise-) simulated image with a Lanczos resampling allows us to better mimic the data reduction process and to better reproduce the background noise, while avoiding an expensive data reduction of simulated raw images.

\begin{figure}
\centering
\includegraphics[width=8cm,angle=0]{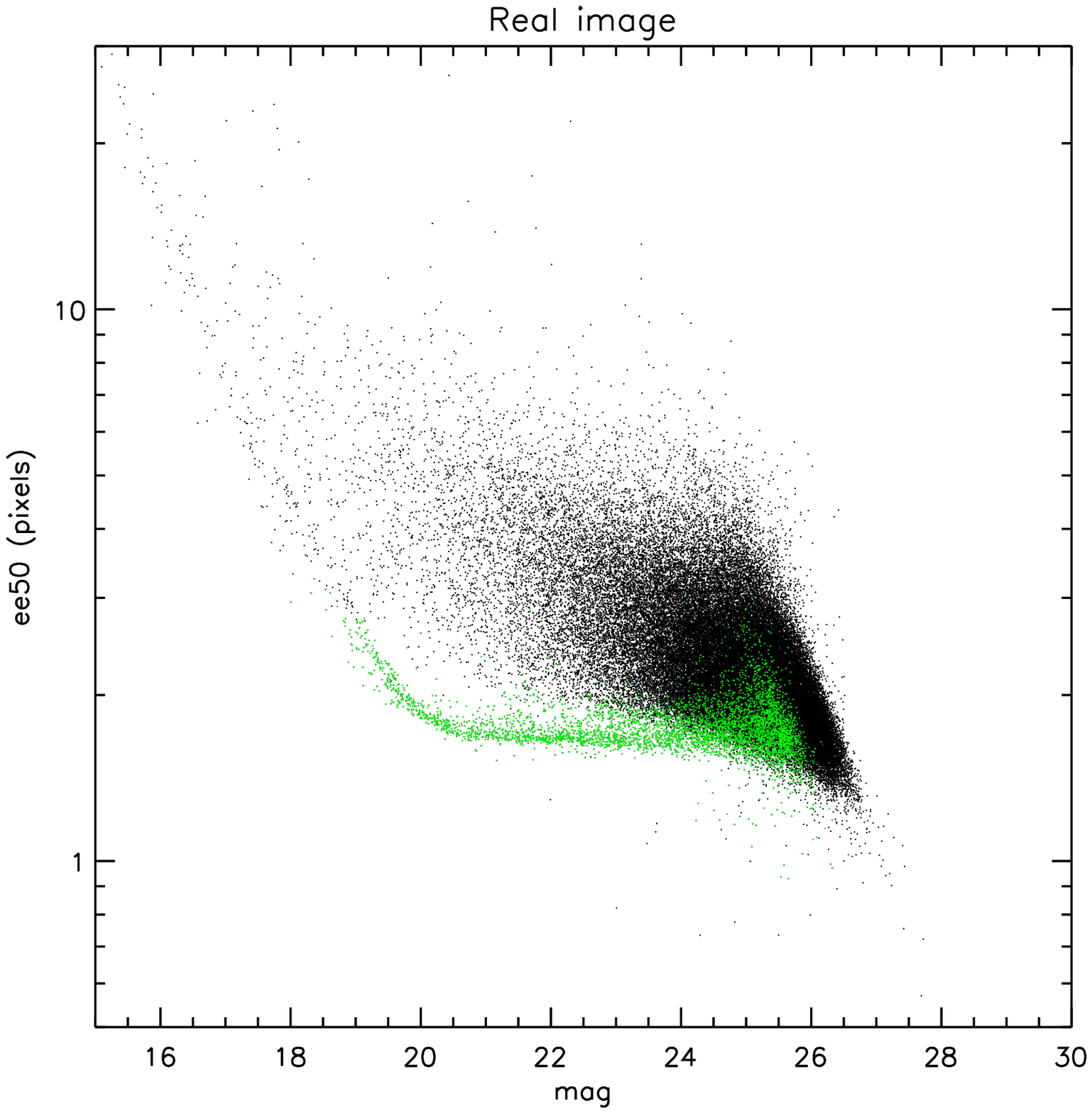}
\includegraphics[width=8cm,angle=0]{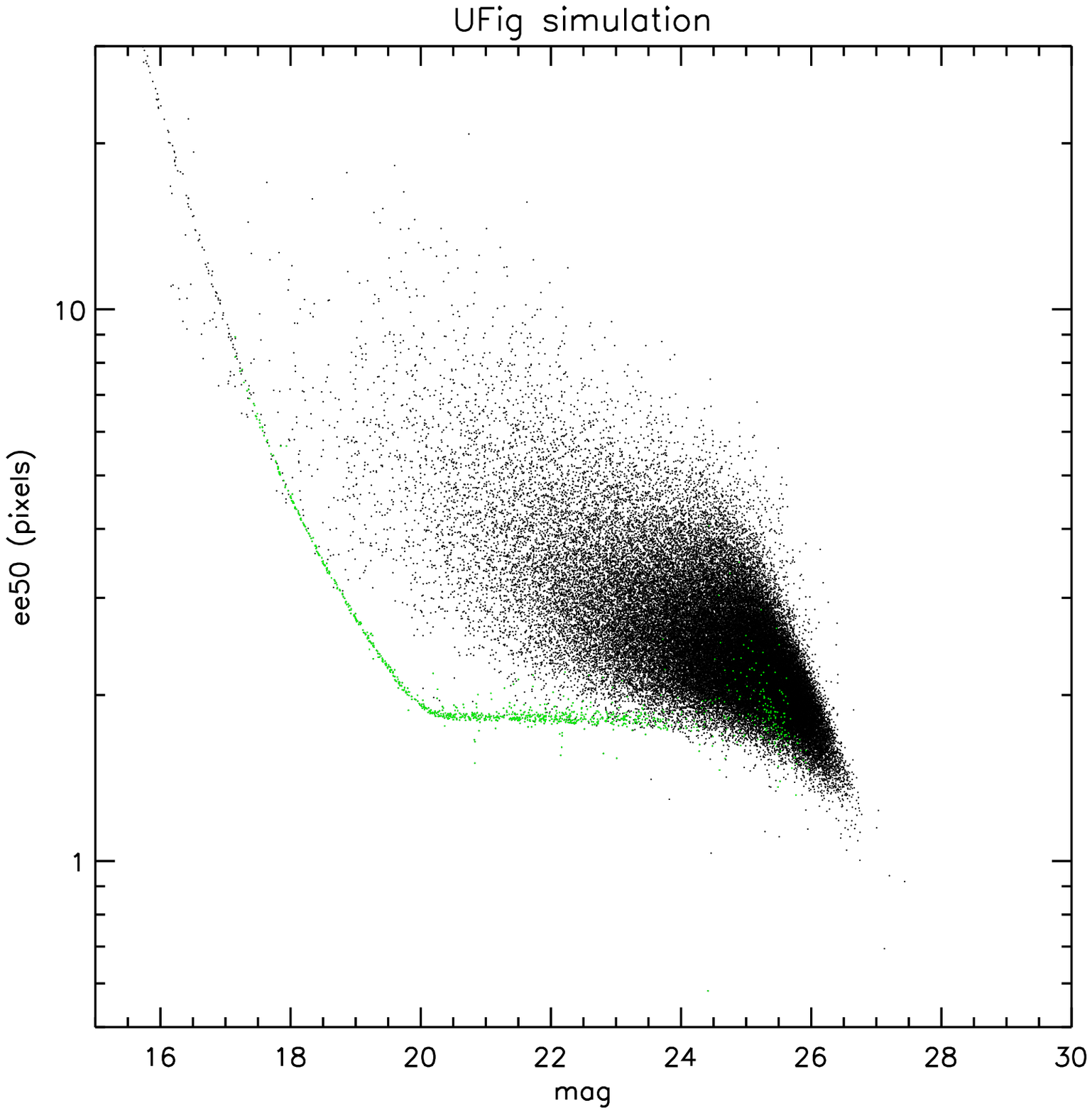}
\caption{\small Magnitude-size distribution for a typical Subaru image (upper panel) and for a {\sc UFig} simulation (lower panel). Black points represent galaxies, and green points represent stars, galaxies and stars being separated by {\sc SExtractor}.} \label{fig_ms}
\end{figure}

Fig. \ref{fig_ms} shows the distribution of stars and galaxies in the magnitude-size plane, both for a typical Subaru image (upper panel) and our simulation (lower panel). Magnitudes are given by {\sc SExtractor}, and we define the half-light radius ee50 as {\sc SExtractor}'s FLUX\_RADIUS (note that all {\sc SExtractor}'s parameters are the same when running {\sc SExtractor} on the real image and on the simulation, preventing any difference from the {\sc SExtractor} analysis).
We also rely on {\sc SExtractor} to perform the galaxy-star separation; to this end, we set {\sc SExtractor}'s SEEING\_FWHM equal to the seeing input in the simulation (0.6'' in the  case shown by Fig. \ref{fig_ms}), and we further set {\sc SExtractor}'s CLASS\_STAR=0.9. Galaxies are shown by the black symbols, while stars are shown by the green symbols.  The star branch is narrower in the simulation than in the real image because we forced the PSF to be constant in the simulation.

Despite the simple models used, {\sc UFig} produces realistic images, that are consistent with real images.

\section{Computational performance} \label{sect_perfs}

\subsection{Execution time}

We tested the performance of {\sc UFig} on a laptop (macBook Pro with a 2.7 GHz Intel processor, 4 cores and 16 GB RAM). Using eight threads, {\sc UFig} provides a 10k$\times$8k-pixels-image (approximately the size of a typical Subaru image) in 30 seconds. This is smaller than the execution time of {\sc SExtractor} on a similar image.
For comparison, it took several hours to create a similar image using {\sc Skymaker} and {\sc Simage}.

We further measured that {\sc UFig} uses 66\% of the laptop's peak performance (29 Gflops out of 43.2 peak Gflops).

\begin{figure}
\centering
\includegraphics[width=8cm,angle=0]{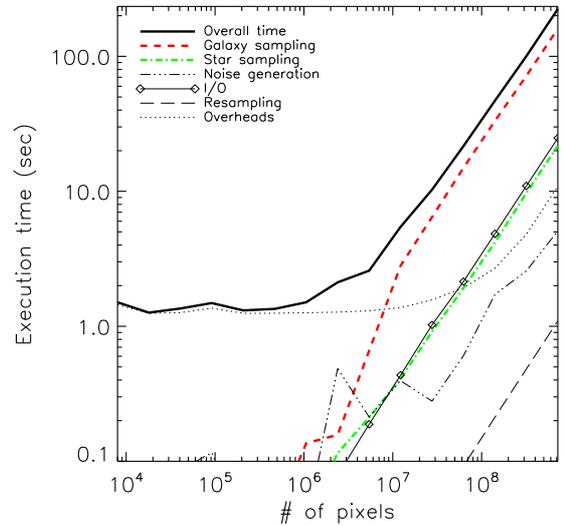}
\caption{\small Execution time vs image size, using eight threads. Black solid line: overall time. Red short dashed line: galaxy sampling. Green dash-dot line: stars sampling. Dash-dot-dot-dot line: noise generation. Diamonds: image and catalog writing to disk. Long dashed line: image resampling. Dotted line: overheads} \label{fig_timsize}
\end{figure}

In the remainder of this section, we discuss how the {\sc UFig} execution time depends on some input parameters, and how the time spending is distributed between different tasks.
Fig. \ref{fig_timsize} shows how the execution time to create an image with exposure time $t_{\rm exp}=450 $s depends on the image's size, when using eight threads. The black thick solid line shows the total execution time. The red thick short dashed line shows the time spent sampling galaxies, while the green thick dash-dot line shows that spent simulating stars. The dash-dot-dot-dot line corresponds to the time needed to generate noise and the long dashed line to the time spent resampling the image. 
Diamonds represent the time spent writing the image and the corresponding catalog to the disk.
Finally, the dotted line shows the overheads (defined as all tasks not directed directly at creating or writing the simulation). 
For big enough images (more than $10^7$ pixels), most of the time is spent drawing galaxies, then drawing stars. For smaller images, the execution time is dominated by overheads. Excepted the overheads, that are expected not to depend strongly on the image size, all tasks' time-spending show a clearly linear dependence on image size. This behavior is expected, since the number of photons (for galaxy generation) and pixels belonging to stars (for star generation) increase linearly with the image size. Similarly, all other tasks, by definition, depend linearly on the number of pixels.

\begin{figure}
\centering
\includegraphics[width=8cm,angle=0]{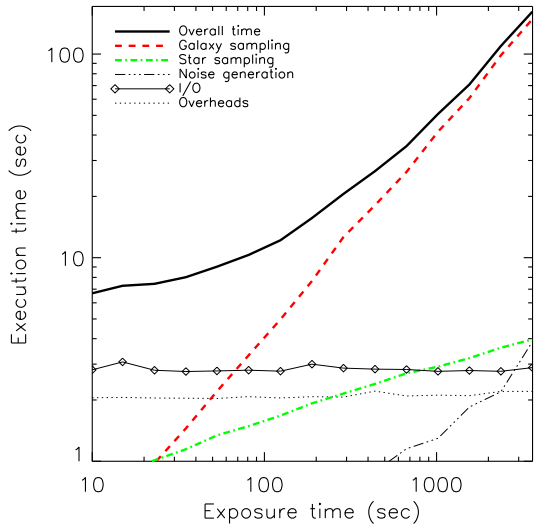}
\caption{\small Execution time vs exposure time, using eight threads. Line styles and colors are the same as in Fig. \ref{fig_timsize}.} \label{fig_texp}
\end{figure}

Fig. \ref{fig_texp} shows the relation between the exposure time used in the simulations, and the execution time, when using eight threads. Lines have the same meaning as in Fig. \ref{fig_timsize}. The time spent resampling the image is not shown, since it is less than one second.
For small exposure times (less than 200 seconds), most of the time is spent writing to disk, whereas generating galaxies is most expensive for large exposure times. Generating galaxies and stars both scale linearly with exposure time. This is expected; the number of photons that one has to simulate to generate galaxies obviously depends linearly on the exposure time.  So does the flux, and therefore the number of pixels that one has to draw when creating stars.
All other tasks do not depend on the exposure time.

\subsection{Parallelization}

\begin{figure}
\centering
\includegraphics[width=8cm,angle=0]{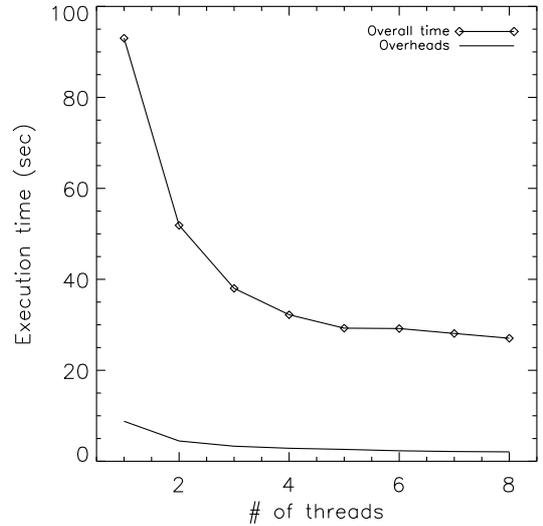}
\caption{\small Execution time vs number of threads. Diamond-solid line: overall execution time. Thin solid line: overheads.} \label{fig_paral}
\end{figure}

Fig. \ref{fig_paral} shows how the execution time depends on the number of threads used for the computation, when simulating a 10,000$\times$8,000 pixels image with an exposure time $t_{\rm exp}=450$ s on the same laptop as that used above. The lower curve corresponds to the overheads, as defined above.
When using two threads, we need half as much time to create a simulation as when using only one thread. A significant gain in execution time can be seen when using up to five threads, before the execution time plateaus. This plateau is due to the fact that Intel hyper-threading is at work, meaning that when more than four cores are used, threads starts to compete against each other (our laptop having four physical cores -- corresponding to eight virtual cores). We therefore expect an optimal parallelization to provide us with a factor of four improvement in execution time: this is indeed what we measure, meaning that our parallelization is nearly optimal.

\subsection{Memory management}

Independently of the number of threads used, a run of the code uses the amount of RAM memory needed to store one copy of the simulated image. For instance, for a 10,000$\times$8,000 pixels simulation, 300 MB of memory are required.

\section{Conclusion and perspective} \label{sect_ccl}

By introducing the Ultra Fast Image Generator ({\sc UFig}), we showed that it is currently possible to implement very fast codes to simulate wide-field astronomical images. We showed that, using simple models, we can simulate realistic images which take observation constraints, including the PSF and various sources of noise, into account. 

Combining analytic simple models with state-of-the-art computing optimizations allows us to produce the mock of a typical Subaru SuprimeCam image (0.25 sq. deg, 10k$\times$8k pixels) with a 5-$\sigma$ (extended) limiting magnitude of ${\rm Rc}=26$ , in which we account for galaxies of magnitude up to ${\rm Rc}=29$, in 30 seconds when using a laptop (macBook Pro with a 2.7 GHz Intel processor, 4 cores and 16 GB RAM); thus, an average galaxy is simulated in 30 $\mu$s.
This represents an improvement of several orders of magnitude in execution time compared to the public softwares that we are aware of. It is also comparable with the execution time of {\sc SExtractor} on a similar image; given the optimization of its implementation, as well as its extensive and intensive use, {\sc SExtractor} can be taken as a standard, well optimized, code in astronomy, thus setting a timescale for any new software that is used in combination with it. 
{\sc UFig} is thus complementary to end-to-end simulation codes which aim to model observational effects in great details, but with greater execution time.

We have presented the implementation of the code, with an emphasis on the different optimizations that we use. In particular, we have found that the optimal solution is to adopt a hybrid approach to generate galaxies and stars, where we create galaxies with a photon-based approach, and stars with a pixel-based approach. We have also described how we optimize random number generation by implementing a lagged Fibonacci random number generator, how we parallelize the code using multithreading in which threads are completely independent, and how we approximate common functions and use SSE vectorization to speed up calculations.

We have then shown that {\sc UFig}'s simulations are consistent with real images, using simple standard tests. Finally, we have investigated the performances of the code, where we have checked that the execution time scales as it should were the code perfectly optimized. In particular, we showed that the parallelization of the code is nearly optimal. Therefore, the {\sc UFig} implementation reaches the limits of the current computation possibilities. This is highlighted by the usage of 66\% of our laptop's peak performance by {\sc UFig}.

The current {\sc UFig} implementation relies on simple models; although it produces images sufficiently realistic for many applications, such as testing data processing codes, or calibrating photometry or astrometry codes, these simple models may need to be refined to use {\sc UFig} for very high-precision analyses. Therefore, we plan to increase the {\sc UFig} realism by including more cosmology in the code. For instance, we will increase the shape complexity of galaxies, allow them to be spatially clustered, as well as distributed in redshift, and we will add weak lensing (either from large-scale structures or from massive clusters). Using more complex models will have an impact on the code's performance, which we will assess and take into account to keep {\sc UFig} optimal. 

Another application of {\sc UFig}'s speed is to calibrate a computer intensive measurement method. 
For instance, \cite{refregier12} and \cite{kacprzak12} showed that a promising approach to cosmic shear measurement pipelines is to calibrate them with image simulations (with observation conditions similar to those of the real data to analyze) to alleviate systematic effects. Up to now, such a calibration was time-consuming and thus limited.
Hence, {\sc UFig} opens a new window to improve on computational intensive measurement techniques, such as those used in weak lensing, or in transient searches, for which {\sc UFig}'s ability to efficiently simulate time series observations may prove to be central.

Further details about {\sc UFig} can be found at http://www.astro.ethz.ch/refregier/research/Software/ufig.

\appendix

\section{Galaxy model} \label{app_gals}

\subsection{Sersic profile}

\begin{figure} 
\includegraphics[width=8cm]{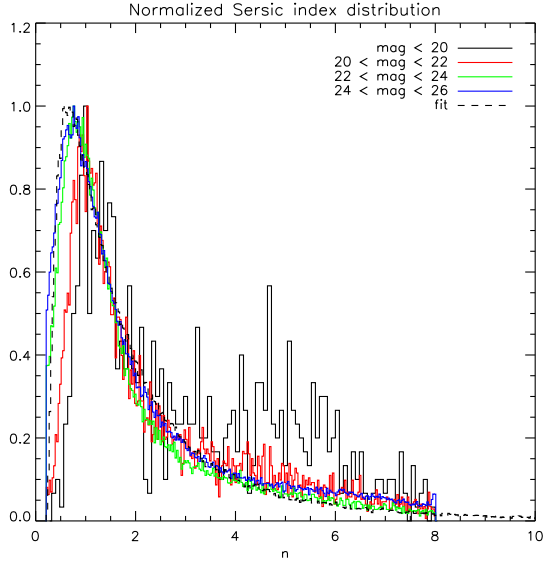}%
\caption{\label{fig_sersic} Distribution of Sersic indices from the ACS-GC catalog \cite{griffith12}, for different magnitude ranges. Histograms are scaled such that their maximum is 1. The dashed line shows the analytical description input in {\sc UFig} for galaxies with magnitude bigger than 20.}
\end{figure}

We describe galaxies by a Sersic profile (Eq. \ref{eq_sersic}).
To assess the distribution of the Sersic index for galaxies up to high redshift, we use the Advanced Camera for Survey General Catalog (ACS-GC -- \cite{griffith12}). Figure \ref{fig_sersic} shows the distribution that we extract from the catalog, as measured in the I-band, for galaxies with good photometric redshifts and magnitude between 15 and 26. Colors code for different ranges of magnitudes: black for magnitude less than 20, red for magnitude between 20 and 22, green for magnitude between 22 and 24, and blue for magnitude between 24 and 26. For bright galaxies, the distribution is clearly bimodal, as is well known (see e.g. \cite{driver06}). This bimodality disappears for fainter galaxies. Whether this highlights physical processes in galaxy formation and evolution, or whether it is simply due to a selection effect, is beyond the scope of this paper: we are only interested in modeling the distribution of Sersic indices as close to the observed one as possible. We find that the p.d.f of Sersic indices for bright galaxies (magnitude less than 20) is well represented by:
\begin{equation}
f(n)=\exp(\mathcal{N}(0.3,0.5)+\mathcal{N}(1.6,0.4))+0.2.
\end{equation}
For faint galaxies (magnitude bigger than 20), we find that the p.d.f of Sersic indices if well described by:
\begin{equation}
f(n)=\exp(\mathcal{N}(0.2,1))+0.2.
\end{equation}
This analytical description is shown for faint galaxies as a dashed line in Fig. \ref{fig_sersic}.

\subsubsection{The circular Sersic profile seen as a $\gamma$-distribution}

The Sersic profile is given by Eq. (\ref{eq_sersic}), and can be normalized to unity flux as:
\begin{equation}
I(r)=\frac{k^{2n}}{2\pi n r_{50}^2 \Gamma(2n)} \exp\left[ -k\left(\frac{r}{r_{50}}\right)^{1/n}\right].
\end{equation}

This profile can be seen as the p.d.f of the radial position of photons from a circular galaxy.
Then let $X$ be the random variable describing the position of a photon. The probability to find the photon in the shell [$X,X+{\rm d}X$] from the center of the galaxy is given by
\begin{equation}
f_X(X){\rm d}X=2\pi X I(X) {\rm d}X.
\end{equation}

Defining the random variable $Y=k\left( \frac{X}{r_{50}}\right)^{1/n}$, it can be shown that it follows a $\gamma$-distribution of shape parameter $2n$:
\begin{equation}
f_Y(Y)=Y^{2n-1} \frac{e^{-Y}}{\Gamma(2n)}.
\end{equation}

\subsubsection{From a circular to an elliptical galaxy}

We define a galaxy's ellipticity ($\varepsilon_1,\varepsilon_2$) through its quadrupoles $J_{ij}$ as $\varepsilon_1+i\varepsilon_2=(J_{11}-J_{22}+2iJ_{12})/(J_{11}+J_{22})$

A circular galaxy can be made elliptical through the transformation $(x_\mathcal{E},y_\mathcal{E})=A(x_\mathcal{C},y_\mathcal{C})$, where $(x_\mathcal{E},y_\mathcal{E})$ and $(x_\mathcal{C},y_\mathcal{C})$ are the photon's coordinates in the elliptical and circular galaxy respectively, and $A$ is the transformation matrix (for $||\varepsilon|| \neq 0$):
\begin{multline}
A=\frac{1}{\sqrt{2}} \times \\
\left(
\begin{matrix}
{\rm sign}(\varepsilon_2) \sqrt{1+||\varepsilon||} \sqrt{1+\frac{\varepsilon_1}{||\varepsilon||}} & -\sqrt{1-||\varepsilon||} \sqrt{1-\frac{\varepsilon_1}{||\varepsilon||}} \\
\sqrt{1+||\varepsilon||} \sqrt{1-\frac{\varepsilon_1}{||\varepsilon||}} & {\rm sign}(\varepsilon_2) \sqrt{1-||\varepsilon||} \sqrt{1+\frac{\varepsilon_1}{||\varepsilon||}}
\end{matrix}
\right)
\end{multline}
If $||\varepsilon||=0$, $A$ is the identity matrix.

\subsection{Magnitude distribution} \label{ssect_galscounts}

\begin{figure} 
\includegraphics[width=8cm]{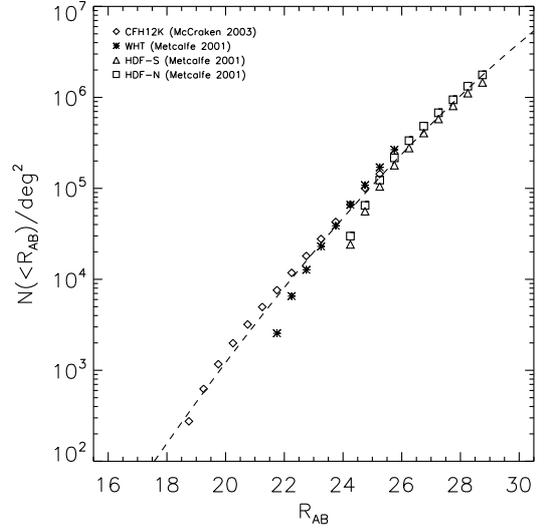}
\caption{\label{fig_galscounts} Cumulative distribution of galaxies' magnitude in the R-band, from several surveys. The fitting function (dashed line) is defined by Eq. (\ref{eq_magdist}) and Table \ref{tab_galscounts}.}
\end{figure}

We parametrize the magnitude distribution of galaxies from galaxy counts in different surveys, such as the VIRMOS Descartes \cite{mccracken03}, COSMOS \cite{capak07}, SXDS \cite{furusawa08}, and the Hershell Telescope and Hubble Deep fields \cite{metcalfe01}. We compile the cumulative counts of these surveys, and fit the resultant overall counts with a polynomial of the form 
\begin{equation} \label{eq_magdist}
\log_{10}(N<{\rm mag})=\sum_i a_i ({\rm mag}-23)^i,
\end{equation} 
as shown by Fig. \ref{fig_galscounts} for counts in the R-band. Table \ref{tab_galscounts} summarizes the coefficients $a_i$ for different filters.

\begin{table}
\caption{\label{tab_galscounts} Fitting coefficients for galaxy cumulative counts.}
\begin{tabular}{cccc}
\hline
\hline
Filter & $a_0$ & $a_1$ & $a_2$ \\
Rc & 4.300 & 0.383 & -0.00766 \\
I & 4.579 & 0.360 & -0.0229 \\
Z & 4.558 & 0.410 & -0.0248 \\
\hline
\end{tabular}
\end{table}

The number of photons making up a galaxy of magnitude {\rm mag}, for a single exposure time $t_{\rm exp}$, in the AB-magnitude system, is given by:
\begin{equation} \label{eq_nphotons}
N_\Phi = 10^{-26} t_{\rm exp} \Delta S \frac{\Delta \lambda}{h \lambda} T 10^{0.4(8.9+\beta {\rm airmass} - {\rm mag})},
\end{equation}
where $\Delta S$ is the effective telescope's mirror's surface, $\lambda$ is the filter's central wavelength, $\Delta \lambda$ the filter's width, $\beta$ is the atmospheric extinction, $h$ is the Planck constant, and $T$ is the total throughput of the observation system $T=T_{\rm mirror} T_{\rm camera} T_{\rm filter} T_{\rm corrector}$.

This number can be rescaled to simulate a galaxy on a coadded image with magnitude zero-point ${\rm mag}_0$:
\begin{equation}
N'_\Phi = N_\Phi 10^{0.4({\rm mag}_0 - {\rm mag}_0^{(I)} - 2.5 \log(t_{\rm exp}))}
\end{equation}
where ${\rm mag}^{(I)}_0$ is the instrument magnitude zero-point, i.e. the magnitude corresponding to a flux of 1 ADU for a one second exposure time in the same observing conditions,
\begin{equation}
{\rm mag}^{(I)}_0 = 8.9 + \beta {\rm airmass} -2.5 \log \frac{{\rm gain}}{Q_E 10^{-26} \Delta S \frac{\Delta \lambda}{h \lambda} T},
\end{equation}
where {\rm gain} is the CCD's gain and $Q_E$ its quantum efficiency.

\subsection{Magnitude-size relation}

We use the ACS-GC catalog to parametrize the relation between the apparent magnitude and apparent size of galaxies. In this catalog, galaxies' $r_{50}$ are estimated by fitting a Sersic profile.
Since this catalog is shallower than the public data we used to estimate the magnitude distribution (\ref{ssect_galscounts}), we do not use it to parameterize the magnitude distribution, but use that derived above.  The ACS-GC data are therefore used only to analytically describe the relation between the magnitude and the size of galaxies. We find that in the ${\rm mag}_r-r_{50,r}$ plane, where ${\rm mag}_r$ and $r_{50,r}$ are the magnitude and the size rotated such that they become uncorrelated, and shifted such that they have zero mean, the distribution of the log of the size $r_{50,r}$ at a given magnitude ${\rm mag}_r$ is well fitted by a Gaussian. Additionally, given that the correlation angle between the size and the magnitude is small, we checked that the magnitude distribution derived in \ref{ssect_galscounts} for the unrotated magnitudes is still a good fit to that of the rotated magnitudes ${\rm mag}_r$.

Thus, we set a galaxy's magnitude and size such that:
\begin{equation}
\left(
\begin{array}{c}
{\rm mag} \\
r_{50} \\
\end{array}
\right)
=
\left(
\begin{array}{cc}
\cos \theta & \sin \theta \\
-\sin \theta & \cos \theta \\
\end{array}
\right)
\left(
\begin{array}{c}
{\rm mag}_r \\
r_{50,r} \\
\end{array}
\right)
+
\left(
\begin{array}{c}
{\rm mag}_p \\
r_{50,p} \\
\end{array}
\right)
\end{equation} 
where ${\rm mag}_p$ and $r_{50,p}$ set the pivot point around which magnitudes and sizes are rotated. They are estimated from the magnitude and size means of COSMOS, and are set to ${\rm mag}_p=25.309$ and $\log r_{50,p}=-0.796 \,\,{\rm arc sec}$. The correlation angle $\theta$ is set to 5.7 deg. The rotated magnitude ${\rm mag}_r$ is drawn from the distribution (\ref{eq_magdist}), and the log of the rotated size $r_{50,r}$ is drawn from a normal distribution of zero mean and standard deviation 0.19 arcsec.

We checked that the magnitude-size relation does not depend significantly on the observing band, and therefore, we restrict its parametrization to what is presented here.

\section{Stars and Point Spread Function models} \label{app_stars}

We use a Moffat profile to account for the PSF.
The (circular) Moffat profile is defined in Eq. (\ref{eq_moffat1}), where $I_0$ is the value at the origin ($r=0$), and $\alpha$ and $\beta$ are scale parameters depending on the observation's conditions. For instance, for realistic atmospheric turbulences, $\beta=4.765$ \cite{trujillo01}. We use $\beta=2.6$ to account for instrumental effects (especially diffraction). The profile's width $\alpha$ is related to its FWHM by $\alpha=\frac{{\rm FWHM}}{2\sqrt{2^{1/\beta}-1}}$ and to its $r_{50}$ by $\alpha=\frac{r_{50}}{\sqrt{2^{1/(\beta-1)}-1}}$.

Contrary to the case of the Sersic profile, the Moffat profile cannot be linked to a usual known distribution, from the p.d.f of which we can easily estimate the displacement to apply to a given photon in {\sc UFig}. Therefore, we sample the displacement by inverting the c.d.f of the profile.

The Moffat profile, seen as the normalized p.d.f $f_X(X)$, where the random variable $X$ corresponds to the photon's displacement, is given by:
\begin{equation} \label{eq_moffat}
f_X(X){\rm d}X = \frac{2(\beta-1)X}{\alpha^2} \left[1+\left(\frac{X}{\alpha}\right)^2\right]^{-\beta}{\rm d}X.
\end{equation}
Defining the variable $Y=1+(X/\alpha)^2$, so that the p.d.f of $Y$ is $f_Y(Y) = (\beta-1)Y^{-\beta}$
for $Y>1$, and integrating it, we obtain the c.d.f of $Y$:
\begin{equation} \label{eq_cdf_moffat}
cdf(Y)=1-Y^{1-\beta}.
\end{equation}

The displacement due to the PSF is then obtained by inverting Eq. (\ref{eq_cdf_moffat}), the c.d.f. itself being uniformally sampled:
\begin{equation}
X=\alpha \sqrt{\left[cdf(Y)-1\right]^{\frac{1}{1-\beta}}-1}.
\end{equation}

We parametrize the stars' magnitude distribution with a polynomial fit to the Milky Way model derived in \cite{robin03}. Given the position in the sky of the image we want to simulate, we extract the stars' magnitude distribution from the corresponding online application \cite{besancon}.

\section{Noise model}

Poisson noise is automatically and naturally accounted for when simulating galaxies in a photon-based approach. We add Poisson noise for stars, and account for the noise from various sources, like the sky brightness, the readout noise and data processing, with a Gaussian random deviate with zero mean and standard deviation (see e.g. \cite{grazian04} or \cite{meneghetti08}):
\begin{equation} \label{eq_fullnoise}
\sigma_N = \sqrt{n_{\rm exp} \left( \frac{RON}{\rm gain}\right)^2 + \frac{F_{\rm sky}}{n_{\rm exp} {\rm gain}} + f_{\rm dp}},
\end{equation}
where $RON$ is the readout noise of the camera, $F_{\rm sky}$ is the sky brightness in ADUs, $n_{\rm exp}$ is the number of exposures out of which the coadded image is assumed to be done and $f_{\rm dp}$ describes the noise coming from the data reduction (including, but not limited to, flat-fielding inaccuracies). 

It should be noted that independently of the hybrid approach we use to simulate galaxies and stars, we treat the background noise at the pixel level, and therefore add it in ADUs to the noiseless image. Finally, we resample our simulations with a Lanczos filter to better mimic the data reduction process.

\section*{Acknowledgements}

We want to thank Stefan M\"uller and Julien Carron for useful discussions, as well as Jason Rhodes and Richard Massey for comments on the manuscript. We acknowledge the usage of the Jet Propulsion Laboratory supercomputers, which are provided by funding from the JPL Office of the Chief Information Officer, and we thank Jason Rhodes for making this usage possible.





\end{document}